\documentclass[11pt,a4paper]{article}
\usepackage{epsfig,graphicx,amsfonts,amsmath}

\usepackage{cite}
\RequirePackage{mathrsfs}
\usepackage{color}

\newlength{\dinwidth}
\newlength{\dinmargin}
\def\eq#1{{Eq.~(\ref{#1})}}
\newcommand{\Le}{\left(}
\newcommand{\Ra}{\right)}

\newcommand{\beq}{\begin{equation}}
\newcommand{\eeq}{\end{equation}}
\newcommand{\beqar}{\begin{eqnarray}}
\newcommand{\eeqar}{\end{eqnarray}}
\newcommand{\ep}{\varepsilon}
\newcommand{\up}{\upsilon}

\newcommand{\tu}{\textsl{u}}
\newcommand{\tv}{\textsl{v}}

\newcommand{\rom}[1]{\uppercase\expandafter{\romannumeral #1\relax}}

\date{}
\begin{document}

\title {{~}\\
{\Large \bf CPTM discrete symmetry, quantum wormholes and cosmological constant problem }}
\author{ 
{~}\\
{\large 
S.~Bondarenko$^{(1) }$
}\\[7mm]
{\it\normalsize  $^{(1) }$ Physics Department, Ariel University, Ariel 40700, Israel}\\
}

\maketitle
\thispagestyle{empty}

\begin{abstract}
 We discuss the consequences of the charge, parity, time and mass (CPTM) extended reversal symmetry for the problems of the vacuum energy density and value of the cosmological constant.
The results obtained are based on the framework with the separation of  extended space-time of the interest on the different regions connected by this symmetry with the 
action of the theory valid for the full space-time and symmetrical with respect to the extended CPTM transformations. The
cosmological constant is arising in the model due the 
gravitational interactions between the different parts of the space-time trough the quantum non-local vertices. It is proposed that the constant's value depends on the form and geometry of the vertices which glue 
the separated parts of the 
extended solution of Einstein equations determining in turn it's classical geometry. The similarity of the proposed model to the bimetric theories of gravitation is also discussed.

\end{abstract}

\section{Introduction}

 In this note we consider the consequences of some extended discrete reversal charge, parity, time and mass  (CPTM) symmetry in application to the 
field theory and general relativity. The proposed symmetry relates the different parts (manifolds) of the extended solution of Einstein equations
preserving the same form of the metric $g$.  
The easiest way to clarify this construction is to consider the different and separated  parts of the extended solution of Einstein equations
defined in the light cone coordinated $u,v$
or corresponding Kruskal-Szekeres coordinates \cite{Kruskal}. 
The extended CPTM transform in this case inverses the sign of these coordinates and
relates the different regions of the extended solution preserving the form of the metric unchanged, 
see \cite{Ser1} for the case of 
Schwarzschild's spacetime and the similar
description of the Reissner-Nordtr\"{o}m space-time in \cite{Frolov} for example. This is a main difference of the proposed transform from the usual CPT one which operates with the quantities defined in the same manifold.
Namely, for the two manifolds, A-manifold and B-manifold for example, with coordinates $x$ and $\tilde{x}$, 
the symmetry $g_{\mu \nu}(x)\,=\,g_{\mu \nu}(\tilde{x})\,=\,\tilde{g}_{\mu \nu}(\tilde{x})$
must be preserved by the extended CPTM symmetry transform:
\beqar\label{Add1}
&\,&
q\,\rightarrow\,-\,\tilde{q}\,,r\,\rightarrow\,-\,\tilde{r}\,,
t\,\rightarrow\,-\,\tilde{t}\,,m_{grav}\,\rightarrow\,-\,\tilde{m}_{grav}\,;\,\,\,
\tilde{q}\,,
\,\tilde{r}\,,
\,\tilde{t}\,,
\,\tilde{m}_{grav}\,>\,0\,; \\
&\,&
CPTM(g_{\mu \nu}(x))\,=\,\tilde{g}_{\mu \nu}(\tilde{x})\,=\,g_{\mu \nu}(\tilde{x})\,.
\eeqar
We underline, see \cite{Ser1}, that the usual radial coordinate is strictly positive and there is a need in an additional B-manifold in order to perform  the \eq{Add1} discrete  P transform, see also next Section. The transformation of the sign of the gravitational mass  in this case can be understood as consequence of the request of the preserving of the symmetry of the metric.
The discussion of the similar construction
in application of the quantum mechanics to the black hole physics can be found also in \cite{Hooft}.

  We consider the framework which consists of different manifolds with the gravitational masses of different signs in each one, see details in \cite{Ser1}.
The general motivation of the introduction of the negative mass in the different cosmological  models is very clear. In any scenario, see \cite{Villata, Chardin} for example, the presence of some kind of repulsive gravitation forces or additional gravitational field in our Universe helps with an explanation of the existence of
dark energy in the models, see also \cite{Petit1,DamKog,Hoss,Kofinas,Fluid,Villata1,Hajd} and references therein.
Therefore, we define the B-manifold  as a part of the extended solution of the Einstein's equations populated by
the negative mass particles. As mentioned above it is also the consequence of the metric's invariance in respect to
the reversion of the sign of the Kruskal-Szekeres coordinates. Although the general form of the metric is not affected 
by the sign of the charge or it's change, the proposed PT transform leads to the charge conjugation as well.
 This result is natural and it allows to resolves the baryon asymmetry problem,
we need the all CPTM transformations in order to obtain overall reversed charge of B-manifold in comparison to A-manifold,
see Section 1 further and discussion in \cite{Ni}.  
Important, that the gravitation properties of the matter of B-manifold after the application of the discrete symmetry
is also described by Einstein equations, see  \cite{Villata} or \cite{Chardin} and \cite{Souriau} for the examples of the 
application of the discrete symmetries in the case of the quantum and classical systems.

 In this formulation the proposed approach can be considered as some version of the Multiverse where, nevertheless, the number of the separated worlds is not arbitrary but defined 
by the type of the extended solution of the
Einstein equations. There are two in the Schwarzschild and infinitely many in the Reissner-Nordstr\"{o}ms extended solution, see for example \cite{Frolov}. Further, for the simplicity, we consider A and B
manifolds only. The generalization for the case of another manifold's number or for the case when exist additional complex  topological structures related by some different transformations,
see \cite{Frolov,Carter,Fuller} for example, 
is straightforward. The theory we consider is not the bimetric one as well, see the examples of the bimetric theories with negative masses introduced
in \cite{Petit1}. Instead, on the classical level, we require an existing of the two non-related equivalent systems of Einstein equations in the each part of the extended solution separately, which due the  symmetry can be written as one system of equations
valid in the full extended space-time in the first approximation. The interaction between the separated system of equations is introduced perturbatively on the next step of the approximation. There are non-local terms which connect the manifolds and which contribute when we begin to account 
the quantum effects. 

 Therefore we consider a connection between the manifolds established by the gravitational force exchange, which are graviton in the case of the weak field approximation.
As mentioned above, the natural candidate for this manifold's gluing is the kind of the foam of vertices which belongs not to the same manifold
but to the two at least. In this case the framework contains two or more manifolds which "talk" each with other by the  non-local  correlators. 
These quantum vertices, similar to some extend to the quantum wormholes, 
have been widely used in the investigation of the cosmological constant problem as well, see \cite{ClassWorm} for example. 
The construction proposed in the note  is a dynamical one, the classical dynamics of each metric of separated manifolds is affected by their mutual interactions even at the absence of the other fields. 
In this case the cosmological constant is arising in the
equations as a result of the mutual influence of the different part of the general manifold by the gravitation forces,
these issues are discussed in the Section 2 and Section 3. In the 
first Section, in turn, we consider the simplest and immediate consequences of the CPTM symmetry for the properties of two quantum scalar fields, $\phi(x)$ and $\tilde{\phi}(\tilde{x})$ defined 
in the two different parts of the extended manifold correspondingly. The Section 4 dedicated to the discussion of the relation and differences of the proposed model with
bimetric models, by construction the frameworks are very similar. The last Section of the note is Conclusion where the main results are summarized and discussed.

\section{Energy density of the vacuum }

  In order to clarify the consequences of the CPTM symmetry we shortly remind results of \cite{Ser1}. Consider the Eddington-Finkelstein coordinates
\beq\label{Sec1001}
\tv\,=\,t\,+\,r^{*}\,=\,t\,+\,r\,+\,2\, M\,\ln\left|\frac{r}{2\,M}\,-\,1\right|\,
\eeq
and
\beq\label{Sec1002}
\tu\,=\,t\,-\,r^{*}\,.
\eeq
The Kruskal-Szekeres coordinates $U,V$, which covers the whole extended space-time, are defined in the different parts of the extended solution.
For example, considering  the region \rom{1} with $r\,>\,2M$ in terms of \cite{Frolov} 
where $U\,<\,0\,,V\,>\,0$ we have: 
\beq\label{Sec1003}
U\,=\,-\,e^{-\tu / 4M}\,,\,\,\,\,V\,=\,e^{\tv / 4M}\,.
\eeq
As demonstrated in \cite{Ser1}, see also \cite{Frolov}, the transition to the separated regions of the solutions can by done by the analytical continuation 
of the coordinates which provided by the 
corresponding change of its signs and reversing of the sign of the gravitational mass. Considering the region \rom{3} 
in definitions of \cite{Frolov}  we obtain:
\beqar\label{Sec8}
U\,=\,-\,e^{-\tu / 4M}\,\rightarrow\,\tilde{U}\,=\,e^{-\tilde{\tu} / 4\tilde{M}}\,=\,-\,U\,,\\
V\,=\,e^{\tv / 4M}\,\rightarrow\,\tilde{V}\,=\,-\,e^{-\tilde{\tv} / 4\tilde{M}}\,=\,-\,V\,.
\eeqar
This inversion of the signs of the $(U,V)$ coordinate axes will hold correspondingly in the all regions of $(U,V)$ plane after 
analytical continuation introduced in \cite{Ser1}.   
Formally, from the point of view of the discrete transform performed in $(U,V)$ plane, the transformations \eq{Sec8}
are equivalent to the full reversion of the Kruskal-Szekeres "time" 
\beq\label{Sec1004}
T\,=\,\frac{1}{2}\,\Le\,V\,+\,U\,\Ra\,\rightarrow\,-\,T
\eeq
and radial "coordinate"
\beq\label{Sec1005}
R\,=\,\frac{1}{2}\,\Le\,V\,-\,U\,\Ra\,\rightarrow\,-\,R
\eeq
in the complete Schwarzschild space-time. Therefore, the introduced $T,R$ coordinates and some
transverse coordinates $X_{\bot}$, all denoted simply as $x$,
further we consider as the correct coordinates in a Fourier transform of the quantum fields.  The corresponding change
in the expressions of the functions after the Fourier transform being formally similar to the conjugation is not the conjugation.
namely, the analytical continuation of the functions from A-manifold to B-manifold 
(CPTM transform)
means the change of the sign of $x$ in corresponding Fourier expressions without it's conjugation as whole.

 For the application of the introduced symmetry we consider A and B manifolds (two Minkowski spaces) as separated parts of the extended solution with non-interacting branches of the scalar quantum field 
defined in each region and
related by the CPTM discrete transform.
Namely, consider the usual quantum scalar field defined in our part (A-manifold) of the extended solution:
\beq\label{Sec1006}
\phi(x)\,=\,\int\,\frac{d^{3}\,k}{(2\pi)^{3/2}\,\sqrt{2\, \omega_{k}}}\,\Le
\phi^{-}(k)\,e^{-\imath\,k\,x}\,+\,\phi^{+}(k)\,e^{\imath\,k\,x}\Ra\,=\, \phi^{*}(x) \,,\,\,\,\,[\phi^{-}(k),\,\phi^{+}(k^{'})]\,=\,\delta^{3}_{k\,k^{'}}\,.
\eeq
The conjugation of the scalar field does not change the expressions, we have simply $(\phi^{-})^{*}\,=\,\phi^{+}$.
In contrast to the conjugation, the CPTM discrete transform acts differently. We have:
\beqar\label{Sec1}
&\,& CPTM(\phi(x))\, = \,CPTM\,\Le\,\int\,\frac{d^{3}\,k}{(2\pi)^{3/2}\,\sqrt{2\,\omega_{k}}}\,\Le
\phi^{-}(k)\,e^{-\imath\,k\,x}\,+\,\phi^{+}(k)(k)\,e^{\imath\,k\,x}\Ra\,\Ra\,\,=\, \tilde{\phi}(\tilde{x})\,=\,
\nonumber \\
&=&\,
\int\,\frac{d^{3}\,k}{(2\pi)^{3/2}\,\sqrt{2\,\tilde{\omega}_{k}}}\,\Le
\phi^{-}(k)\,e^{\imath\,k\,\tilde{x}}\,+\,\phi^{+}(k)(k)\,e^{-\imath\,k\,\tilde{x}}\Ra\,=\, 
\nonumber \\
&=&\,
\int\,\frac{d^{3}\,k}{(2\pi)^{3/2}\,\sqrt{2\,\tilde{\omega}_{k}}}\,\Le \tilde{\phi}^{+}(k)\,e^{\imath\,k\,\tilde{x}}\,+\,
\tilde{\phi}^{-}(k)\,e^{-\imath\,k\,\tilde{x}}\,\Ra\,
\eeqar
that provides:
\beq\label{Sec2}
\phi^{-}(k)\,\rightarrow\,\tilde{\phi}^{+}(k)\,,\,\,\,\phi^{+}(k)\,\rightarrow\,\tilde{\phi}^{-}(k)\,,\,\,\,\omega(k)\,=\,\sqrt{m^2 + k^2}\rightarrow\,\omega(k)\,=\,\sqrt{m^2 + k^2}\,.
\eeq
Now we face a problem. Indeed, using usual commutator's definition, we have to write
\beq\label{Sec2001}
[\tilde{\phi}^{-}(k),\tilde{\phi}^{+}(k^{'})]\,=\,\delta^{3}_{k\,k^{'}}\,=\,[\phi^{+}(k),\,\phi^{-}(k^{'})]\,
\eeq
that contradicts to the \eq{Sec1006} commutation relations. Therefore we redefine the commutator of the new operators:
\beq\label{Sec201}
[\tilde{\phi}^{-}(k),\tilde{\phi}^{+}(k^{'})]\,=\,-\,\delta^{3}_{k\,k^{'}}\,,
\eeq
this is the difference of the conjugation of the field and continuation of the field to the another region of the extended manifold. 
The consequence of this new commutation relation is simple.
We write the general energy-momentum vector written for the both regions of the manifold
\beq\label{Sec4}
P^{\mu}\,=\,\frac{1}{2}\,\int\,d^{3}\,k\,k^{\mu}\,\Le \phi^{+}(k)\,\phi^{-}(k)\,+\,\phi^{-}(k)\,\phi^{+}(k)\,+\,
\tilde{\phi}^{+}(k)\,\tilde{\phi}^{-}(k)\,+\,\tilde{\phi}^{-}(k)\,\tilde{\phi}^{+}(k)\,\Ra\,
\eeq
and using the commutators of the two sets of the operators, \eq{Sec1006} and \eq{Sec201}, we obtain
\beq\label{Sec41}
P^{\mu}\,=\,\int\,d^{3}\,k\,k^{\mu}\,\Le \phi^{+}(k)\,\phi^{-}(k)\,+
\tilde{\phi}^{+}(k)\,\tilde{\phi}^{-}(k)\,\Ra\,=\,P^{\mu}_{1}\,+\,P^{\mu}_{2}\,
\eeq
with precise cancellation of the vacuum zero modes contributions, here $P^{\mu}_{i}$ are energy-momentum vectors defined in A-manifold and B-manifold separately.
The same holds as well for the case of charged scalar field where
additionally the Ctransform provide
for the mutual charge operator of the both parts of the solution:
\beq\label{Sec6}
Q\propto\,a^{*}\,a\,-\,b^{*}\,b\,+\,\tilde{a}^{*}\,\tilde{a}\,-\,\tilde{b}^{*}\,\tilde{b}\,=\,0\,
\eeq
as we expect for the overall charge of the regions related by the discrete C transform. 
Therefore, the result of the introduced symmetry is that the vacuum energy density is precisely zero on the classical level when we consider two non-interacting branches of the quantum field related by the CPTM transform, see discussion in \cite{Wienberg}.

\section{Cosmological constant through the gravity's modified action}

 Our next step is an introduction of the two regions of the full space-time connected by the extended CPTM symmetry with the possible presence of the scalar fields separately in the each region.
We introduce the partition function which preserves the symmetry discussed above which relates the two separated parts of the space-time:
\beq\label{Sc1}
Z\,=\,Z_{0}^{-1}\,\int\,D g_{\mu \nu}\,D \phi(x)\,D \tilde{\phi}(\tilde{x})\,e^{\imath\,S[g,\,\phi(x),\,\tilde{\phi}(\tilde{x})]}
\eeq
with ($c\,=\,\hbar\,=\,1$)
\beq\label{Sc2}
S\,=\,-\,\frac{1}{16\,\pi\,G}\,\int\,d\Omega\,\sqrt{-g}\,R\,+\,\int\,d^4 x\,\sqrt{-g}\,{\cal L}(\phi(x))\,+\,\int\,d^4 \tilde{x}\,\sqrt{-g}\,{\cal L}(\tilde{\phi}(\tilde{x}))\,-\,
S_{int}(g,\,\phi(x),\,\tilde{\phi}(\tilde{x}))\,.
\eeq
Here the gravitational filed is defined everywhere in the space-time related by the CPTM transform,  i.e.
\beq\label{Sc3}
d\Omega\,\sqrt{-g}\,=\,d^4 x\,\sqrt{-g(x)}\,+\,d^4 \tilde{x}\sqrt{-g(\tilde{x})}\,\,,
\eeq
whereas for the scalar fields we wrote the Lagrangians  separately in the each region due the difference in the commutation relations and 
consequent difference of the corresponding Green's functions.
Now we can try to guess the possible form of the interaction part in the \eq{Sc2} action. We request that this term will preserve the deserved symmetry  
of the problem and that the interaction between the fields is carried out only through the gravity, i.e. by the fluctuations around any classical metric. The simplest 
possible variant of the interaction term has the form of the sum of source terms for the fields which must be non-local in this case, see \cite{DamKog} for the similar set-up in the case of local interaction term.
For the case of the
scalar field we define in the A and B manifolds separately:
\beq\label{Sc4}
S_{\phi\,int}=\int\,d^4 x\,\sqrt{-g(x)}\,\int d^4 \tilde{x}\sqrt{-g(\tilde{x})}\,\xi_{\phi}(\tilde{x}, x)\,\phi(x) +
\int d^4 \tilde{x}\sqrt{-g(\tilde{x})}\,\int\,d^4 x\,\sqrt{-g(x)}\,\xi_{\tilde{\phi}}(x, \tilde{x})\,\tilde{\phi}(\tilde{x})\,,
\eeq
see \cite{Linde} as well 
\footnote{We also can assume the version of the interaction term with $\phi\,\rightarrow\,\phi^{n}$ change and corresponding redefinition of the the dimension of $\xi$ function, but we consider the expression as the simplest type of the source term preserving $n\,=\,1$ value.} . 
Additional interaction term in \eq{Sc2} we introduce for the pure gravitational interactions between the manifolds, it's
simplest version can be written as:
\beq\label{Sc5}
S_{g\,int}=\int\,d^4 x\,\sqrt{-g(x)}\,\int d^4 \tilde{x}\sqrt{-g(\tilde{x})}\,\xi_{g}(x, \tilde{x})\,.
\eeq
Now consider the case without matter fields present. We obtain for the \eq{Sc1}:
\beq\label{Sc6}
Z_{g}\,=\,Z_{0\,g}^{-1}\,\int\,D g_{\mu \nu}(x)\,D g_{\mu \nu}(\tilde{x})\,e^{\imath\,S_{g}[g(x),\, g(\tilde{x})]}\,
\eeq
with pure gravitational action
\beq\label{Sc7}
S_{g}=-\,\frac{1}{16\,\pi\,G}\,\int\,d^4 x\,\sqrt{-g}\,R\,-\,\frac{1}{16\,\pi\,G}\,\int\,d^4 \tilde{x}\,\sqrt{-g}\,R\,-\,
\int\,d^4 x\,\sqrt{-g(x)}\,\int d^4 \tilde{x}\sqrt{-g(\tilde{x})}\,\xi_{g}(x, \tilde{x})\,.
\eeq
The equations of motion for the gravitational field in the each region have the same form and  look as 
\beq\label{Sc101}
R_{\mu \nu}\,-\,\frac{1}{2}\,g_{\mu \nu}\,R\,-\,8 \pi G\,g_{\mu \nu}\,\int d^4 \tilde{x}\sqrt{-g(\tilde{x})}\,\xi_{g}(x, \tilde{x})\,=\,0\,
\eeq
plus the equation with $x\,\rightarrow\,\tilde{x}$ replace. The equations provide the "matter" terms in the expressions even in the case of absent of the real matter,
but the role of the $\xi$ function is still unclear here.
So, further we perform an integration (averaging) with respect to $g(\tilde{x})$ metric, and obtain a modified partition function averaged over the second part of the full space-time:
\beq\label{Sc8}
\bar{Z}_{g}\,=\,\bar{Z}_{0\,g}^{-1}\,\int\,D g_{\mu \nu}(x)\,e^{\imath\,\bar{S}_{g}[g(x)]}
\eeq
where
\beq\label{Sc9}
\bar{S}_{g}[g(x)]\,=\,-\,\frac{1}{16\,\pi\,G}\,\int\,d^4 x\,\sqrt{-g}\,R\,-\,\int\,d^4 x\,\sqrt{-g}\,<\xi_{g}(x)>\,.
\eeq
Here, as usual, $<\xi_{g}(x)>$ means the averaging of the interaction filed with respect to $g(\tilde{x})$, the bare effective gravitational action $\Gamma[g(\tilde{x})]$ is canceled here by 
the corresponding $Z_{0}^{-1}$ 
constant. The resulting equations of motion read as 
\beq\label{Sc10}
R_{\mu \nu}\,-\,\frac{1}{2}\,g_{\mu \nu}\,R\,-\,8 \pi G\,g_{\mu \nu}\,<\xi_{g}(x)>\,=\,0\,.
\eeq
Now the \eq{Sc10} has a familiar form,
the introduced term can be considered as a density of the vacuum energy:
\beq\label{Sc102}
<\xi_{g}(x)>\,\propto\,\rho_{vac}\,
\eeq
which is equal to zero at the classical level, see \eq{Sec41} and next Section.
Identifying  this contribution with the cosmological constant
\beq\label{Sc11}
\Lambda_{g}\,=\,8 \pi G\,<\xi_{g}>\,=\,const\,,
\eeq
we also conclude that the constant is a dynamical variable which depends on the overall evolution of the manifolds and which value in principle can be changed. 
Namely, assuming that the CPTM symmetry is precise at every moment of the evolution and that the classical value of the constant is always zero, we can not say a lot about the 
contribution of the quantum effects during the manifold's evolution. The quantum contributions can be quite large at the different moments of the time depending, 
perhaps, on the curvature of the manifolds. 
Therefore, we discuss a smallness of the constant at the present moment of the A and B manifolds global times when the expansion around a flat space-time is justified. 
Further we will see that it is small due it's non-classical origin. Namely, it is zero at classical level for the flat manifolds and  it's non-zero value at the present   
is due the  quantum corrections to the classical result. These corrections, in turn, depend on the form of the quantum propagators in some auxiliary theory,  
see next section for the short discussion of the issue. 

 Now we can add to the \eq{Sc4} the part of the action responsible for the interaction of the gravitation field with the scalar ones. In this case we can subsequently average the corresponding parts of 
\eq{Sc2} with respect to the $\tilde{\phi}$ and $g(\tilde{x})$ fields. The partition function will acquire the following form therefore:
\beq\label{Sc16}
Z_{g\phi}\,=\,Z_{0\,\phi}^{-1}\,\int\,D g_{\mu \nu}(x)\,D \phi(x)\,e^{\imath\,S_{g\phi}[g(x),\, \phi(x)]}\,
\eeq
with the action after the averaging:
\beqar\label{Sc17}
S_{g\phi} &= & -\,\frac{1}{16\,\pi\,G}\,\int\,d^4 x\,\sqrt{-g}\,R\,+\,
\int\,d^4 x\,\sqrt{-g}\,{\cal L}(\phi(x))\,-\,
\int\,d^4 x\,\sqrt{-g}\,<\xi_{g}(g(x),\,\phi(x))>\,-\,\nonumber \\
&-&
\int\,d^4 x\,\sqrt{-g}\,<\xi_{\phi}(g(x),\,\phi(x))>\,.
\eeqar
The contributions of the last two terms in \eq{Sc17} we can combine into the joint energy-momentum tensor writing the equations of motion as
\beq\label{Sc18}
R_{\mu \nu}\,-\,\frac{1}{2}\,g_{\mu \nu}\,R\,-\,8 \pi G\,T_{g\,\mu \nu}\,=\,8 \pi G\,T_{\phi\,\mu \nu}\,,
\eeq
which reproduce the \eq{Sc10} at the limit of zero scalar field. The energy-momentum tensor in r.h.s. of \eq{Sc18}, in turn, is provided by the $\tilde{\phi}$ field in an another part of the extended solution
which affects on the $\phi$ field in our part of the space-time only through the gravitational interactions, it's action therefore is similar to the dark matter effect. 

 We have to note also\footnote{We grateful to the referee which point out this issue.}, that in general the addition of the scalar fields in the problem leads to the 
additional contributions to the value of the cosmological constant. With the classical contribution of the scalar fields to the constant equal to zero, see Section 1, we notice that 
there are also quantum contributions 
to the energy-momentum tensor comparable with the classical ones. These contributions are proportional to $m^4$ of the fields, see for example \cite{CcQ} and references there in. 
So far we do not know the resolution of this particular problem. For example, there is a possibility of the 
cancellation of such contributions in the framework, similarly to done in \cite{Sym}, where due the opposite signs of the contributions they are canceled.

\section{Quantum vertices of the model}

 In the previous section we did  not specify how to derive the $\xi_{g}$ and $\xi_{\phi}$ functions, 
the only assumption there made was about their non-classical origin and their zero value at the classical level. 
This condition must be satisfied not only at the case of the flat Minkowski space but also at the case of an 
arbitrary topology of the manifolds simply by request of CPTM symmetry and request of the preserving of the form of classical Einstein equations.
Namely, the introduced functions describe the  non-local interactions between A and B manifolds, we do not change the usual form of the free classical gravity action. Such non-local vertices are arising in the description of quantum interaction effects, see
\cite{Foam,LipatovG,Our}.
Therefore, 
we propose to consider the interactions terms as some quantum wormholes with external "legs" placed on the separated manifolds, similarly to the wormholes 
of \cite{Foam,Thorne}. 
In our case the vertices connect not only separate parts of the same manifold but glues different manifolds of the extended solution as well,
they are  kind of wormholes in the Multiverse universe.
In general as example of the
vertices we can consider  the quantum foam of wormholes of \cite{Foam}. For they construction we consider the fluctuations of the gravitational field  around 
some classic solution
\beq\label{Worm1}
\,g_{\mu \nu}\,=\,g_{0\,\mu \nu}\,+\,h_{\mu \nu}(x)\,;\,\,\,\,
\,\tilde{g}_{\mu \nu}\,=\,\tilde{g}_{0\,\mu \nu}\,+\,\tilde{h}_{\mu \nu}(\tilde{x})\,,\,\,\,
CPTM(g_{\mu \nu}(x))\,=\,\tilde{g}_{\mu \nu}(\tilde{x})\,=\,g_{\mu \nu}(\tilde{x})\,
\eeq 
and then we can calculate an effective action of the following form which describes the interaction between the two 
manifolds:
\beqar\label{Worm101}
&\,&\Gamma_{w}\,=\, \nonumber \\
&=&
\sum_{l,\,k=\ 1}\,h_{\mu_1 \nu_1}(x)\,\cdots\,h_{\mu_k \nu_k}(x)\,V^{\mu_{1} \nu_{1}\cdots\mu_{k} \nu_{k};\,\rho_1 \sigma_1\,\cdots\,\rho_l \sigma_l }_{k l}
(x_{1}\cdots\,x_{k};\tilde{x}_{1}\cdots\,\tilde{x}_{l})\tilde{h}_{\rho_1 \sigma_1}(\tilde{x})\,\cdots\,\tilde{h}_{\rho_l \sigma_l}(\tilde{x})\,+\,\nonumber \\
&+&\,
\,h_{\mu \nu}(x)\,V^{\mu \nu}_{1\, 0}\,+\,V^{\rho \sigma}_{0\, 1}\,\tilde{h}_{\rho \sigma}(\tilde{x})\,
\eeqar
with $V_{k\,l}$ as effective vertices of the theory which connect the different manifolds. 
The last two terms in the expression are sources of the gravitation, at the absence of the matter these terms are equal to zero
and in this case the $h,\,\tilde{h}$ fields are quantum fluctuations above the classical solutions. 
The proposed effective action is similar, to some extend, to the effective actions of
\cite{LipatovG,Our}, see there the examples of  the calculations. 

 In order to demonstrate the smallness of the cosmological constant value in the formalism and the way how the vertices of the \eq{Worm101} can arise in the approach, 
let's make an usual exercise and introduce some auxiliary fields in the problem. Namely,
let's rewrite the simplest non-local vertex as following:
\beqar\label{DopE1}
&\,& e^{-m^{6}_{p}\,\int\,d^4 x\,\int d^4 \tilde{x}\,\sqrt{-g(x)}\,\xi_{g}(x, \tilde{x})\,\sqrt{-g(\tilde{x})}} \propto
\nonumber \\
&\propto&\,
\int D\upsilon(x) D \tilde{\upsilon}(\tilde{x}) \, e^{ \frac{1}{4} \int d^4 x \, \int d^4 \tilde{x} \,\upsilon(x) \hat{A}(x,\tilde{x}) \tilde{\upsilon}(\tilde{x}) - 
\,\int d^4 x\,\upsilon(x)\,J_{1}(x)
- \,\int d^4 \tilde{x}\,J_{2}(\tilde{x})\,\tilde{\upsilon}(\tilde{x}) }
\eeqar
with $m_{p}$ as Planck mass introduced as a regulator of the dimensions  and some operator $\hat{A}(x)$.
Now we have to propose some model for the fields $\upsilon$ and $\tilde{\upsilon}$. The usual interpretation of these fields as the scalar ones is not acceptable here, we
request zero contributions of the fields to the value of cosmological constant. The first example we consider is a definition of the $\upsilon$ and $\tilde{\upsilon}$ fields as doublets
of scalar fields on the A,B manifolds related by the CPTM symmetry. We define:
\beq\label{DopE111}
\up \, = \,\Le\, \Phi_{1}(x)\,\tilde{\Phi}_{1}(\tilde{x})\,\Ra\,,\,\,\,
\tilde{\up}\,=\Le
\begin{array}{c}
\Phi_2(y) \\
\tilde{\Phi}_{2}(\tilde{y})
\end{array} 
\Ra\,,\,\,\,\,
\hat{A}(x,\tilde{x}) = \Le
\begin{array}{c c}
D^{-1}_{1 1}(x, y) & 0 \\
0 & D^{-1}_{2 2}(\tilde{x}, \tilde{y}) 
\end{array} 
\Ra\,.
\eeq
Here $D_{i j}$ are some propagators of the scalar fields and defining the external currents 
\beq\label{DopE112}
J_1\, = \, m^{3}_{p}\, \Le
\begin{array}{c}
\sqrt{-g}\, \\
\sqrt{-\tilde{g}}\,
\end{array} 
\Ra\,,\,\,\,\,
J_2\,=\,m^{3}_{p}\,\Le\, \sqrt{-g}\,\,\sqrt{-\tilde{g}}\,\Ra\,,\,\,\,
\eeq
we rewrite the r.h.s of \eq{DopE1} as
\beq\label{DopE113}
\int D\Phi_{i} D \tilde{\Phi}_{i} e^{\Le \frac{1}{4} \int d^4 x  \int d^4 y \, \Phi_{1} D^{-1}_{1 1} \Phi_{2}  + 
\frac{1}{4} \int d^4 \tilde{x}  \int d^4 \tilde{y} \, \tilde{\Phi}_{1} D^{-1}_{2 2} \tilde{\Phi}_{2}
- m_{p}^{3}\,\sum \Le \int d^4 x\,\sqrt{-g}\, \Phi_{i} +  
\int d^4 \tilde{x}\,\sqrt{-\tilde{g}}\,\tilde{\Phi}_{i} \Ra\,\Ra }\,.
\eeq
Now we have two different scalar fields on the separated manifolds and due the results of Section 1 their contributions to the cosmological constant is zero.
In turn, now we can estimate the form of the $\xi_{g}$ function:
\beq\label{Worm2}
\xi_{g}(x,\,\tilde{x})\,\propto\,<\,\Phi(x)\,\Phi(\tilde{x})\,>\,+\,<\,\tilde{\Phi}(\tilde{x})\,\tilde{\Phi}(x)\,>\,\propto\,D_{11}(x,\,\tilde{x},\,g_{\mu \nu})\,+\,
D_{22}(\tilde{x},\,x,\,\tilde{g}_{\mu \nu})\,,
\eeq
here we  applied the CPTM transform for the $y,\,\tilde{y}$ variables of integration, used the $CPTM(\sqrt{-g})\,=\,\sqrt{-\tilde{g}}$ equality
and extend the  integration over $x$ and $\tilde{x}$ 
coordinates in the expressions over both manifolds in each separated integral\footnote{This extension requires redefinition of the corresponding propagators in the expressions, they are not invariant 
with respect to the symmetry transform. Nevertheless, we require that after all the first terms in the propagator's expansion with respect to the space-time curvature will be scalar propagator in the flat space-time.}.
To the leading order approximation we keep in \eq{Worm2} only the first term of the expansion of the $\sqrt{-g}$ around some classical metric of interests
and preserve only first, flat terms in the expressions for the $D$ propagators in the curve space-time. Defining
$D_{0}(x,\,\tilde{x},\,g_{0\,\mu \nu}^{w} )$ as the propagator of the usual scalar field,  we obtain after the proper regularization:
\beq\label{Worm22}
\Lambda\,\propto\,<\xi_{g}(x)>\,\propto\,\int\,d^{4} \tilde{x}\, D_{0}(x,\,\tilde{x},\,g_{0\,\mu \nu}^{w} ) \,\propto\,
\int\,\frac{d^{4} \tilde{x}}{(x\,-\,\tilde{x})^2}\,=\,0\,,
\eeq
i.e. non-zero value of  $\Lambda$ in pure gravity is possible only due the higher orders of the $\sqrt{-g}$ expansion in respect to some fluctuations around classical solution or
due the higher order of the 
propagator's expansion in the curved space time. This is the reason why  the cosmological constant in the model must be small, it has pure quantum origin.
In general, there exist also more complicated quantum vertices which glue the manifolds and which arise if we will introduce some interactions between $\up$ and $\tilde{\up}$ fields.

 The origin of these complex vertices can be clarified if we will introduce the interactions between the \eq{DopE1} auxiliary fields. 
Namely, consider the following action for the fields:
\beq\label{DopE3}
S\,\propto\,\frac{1}{4} \int d^4 x \, \int d^4 \tilde{x} \,\upsilon(x) \hat{A}(x,\tilde{x}) \tilde{\upsilon}(\tilde{x}) - 
\,\int d^4 x\,\upsilon\,J_{1}(x)
- \,\int d^4 \tilde{x}\,J_{2}(\tilde{x})\,\tilde{\upsilon}(\tilde{x})
 + \int d^4 x \,\int d^4 \tilde{x} \,V(\up,\tilde{\up})\,
\eeq 
with potential $V(\up,\tilde{\up})$ which is defined so that it will provide to the leading order approximation for the one or for the both fields in each doublet:
\beq\label{DopE4}
\Phi_{cl, i}(x)\,=\,m_{p}\,\sqrt{-g}\,,\,\,\,\,\tilde{\Phi}_{cl, i}(\tilde{x})\,=\,m_{p}\,\sqrt{-\tilde{g}}\,.
\eeq
Then we can construct the quantum effective action of interests by expanding the auxiliary fields around the classical background:
 \beq\label{DopE5}
\Phi_{i}(x)\,=\Psi_{cl, i}(x)\,+\,\xi_{i}\,,\,\,\,\,\tilde{\Phi}_{i}(\tilde{x})\,=\tilde{\Psi}_{cl,i}(\tilde{x})\,+\,\tilde{\xi}_{i}\,.
\eeq
Integrating out the fluctuations and expanding the expressions with respect to $\Psi_{cl}(x)$ and $\tilde{\Psi}_{cl}(\tilde{x})$
we  will obtain the  auxiliary effective action of the following form:
\beqar\label{Worm4}
\Gamma_{\Phi\,\tilde{\Phi}}\,& \propto &\,\sum_{k,\,l}\int\,d^4 x\,\sqrt{-g(x_1)}\,\int d^4 x_l\sqrt{-g(x_k)}\,\cdots\,
\int d^4 \tilde{x}_1\sqrt{-g(\tilde{x}_1)}\,
\nonumber \\
&\,&
\int d^4 \tilde{x}_k\sqrt{-g(\tilde{x}_l)}\, \xi_{g}(x_{1},\,\cdots\,,x_{k},\,\tilde{x}_{1},\,\cdots\,,\tilde{x}_{l}; g,\,\tilde{g})\,
\eeqar
with many-legs vertices of interest. 
There are few important points we need to clarify. The first one is that the non-local vertices we consider are the vertices of the interactions between the traces of fluctuations around the classical geometries of  two (or more) separated manifolds, they are quantum. 
The second point is that there are other additional contributions to the effective action which describes the quantum interactions through the wormholes which belong to each manifold separately,
these are quantum wormholes which connect the separated parts of the same manifold.
Also
it must be underlined, that the effective actions \eq{Worm101} and \eq{Worm4} are the different ones. The second one is the effective action of the auxiliary fields which provide
the effective vertices in the classical action \eq{Sc2}. After that, on the base of \eq{Sc7} classical action, the pure gravity effective action \eq{Worm101} can be calculated.
This two-layer calculation procedure is complicated of course but also interesting from the point of view of the renormalization of the theory. There is  a request of the finiteness of the whole 
theory which consists of two parts, each from the separated effective actions can be non-renormalizable in principle.
The last comment about the \eq{Worm4} expression is that the quantum vertices depend on the metrics of the manifolds. It means that they contribute in the equations of motion as well, they are not scalars anymore. Therefore, as we mentioned above, what we defined as cosmological constant in the approach is a sum of complex and small quantum non-constant contributions into the classical 
Einstein equations. Their definitions as cosmological constant requires some formalization and definitely a renormalization.

 An another interesting example of the model for the $\upsilon$ and $\tilde{\upsilon}$ fields is coming from the non-equilibrium condensed matter physics. 
Following to the Keldysh formalism at $T=0$, see \cite{Kamenev}, we can write:
\beq\label{Worm5}
\up\, = \,\Le\, \Phi_{cl}(x)\,\ep(x)\,\Ra\,,\,\,\,
\tilde{\up}\,=\,
\Le
\begin{array}{c}
\tilde{\Phi}_{cl}(\tilde{x}) \\
\tilde{\ep}(\tilde{x})
\end{array} 
\Ra\,,\,\,\,\,
\hat{A}(x,\tilde{x}) = \Le
\begin{array}{c c}
0 & m_{p}^{2}\,\Le D^{A} \Ra^{-1} \\
m_{p}^{2}\,\Le D^{R} \Ra^{-1} & \Le D^{K} \Ra^{-1} 
\end{array} 
\Ra\,.
\eeq
Here $D$ are retarded, advanced and Keldysh propagators correspondingly, $\Phi_{cl}$ and $\ep$ are classical and quantum fields of interests.
In this case the r.h.s. of \eq{DopE1} can be written through the following action:
\beq\label{Worm6}
S\,\propto\,\frac{1}{4}\,\int\,d^4 x\, d^{4}  \tilde{x}\,\Le\,\up(x)\,\hat{A}(x,\,\tilde{x})\,\tilde{\up}(\tilde{x})\,\Ra\,.
\eeq
Now we define:
\beq\label{Worm7}
\Phi_{cl}(x)\,=\,-\,4\,m_{p}\,D^{A}(x,\,\tilde{x})\,\sqrt{-\tilde{g}}\,,\,\,\,\tilde{\Phi}_{cl}(\tilde{x})\,=\,-\,4\,m_{p}\,D^{R}(\tilde{x},\,x)\,\sqrt{-g}\,
\eeq 
and obtain for the \eq{Worm6}:
\beq\label{Worm8}
S\,\propto\,
\frac{1}{4} \int d^4 x \, \int d^4 \tilde{x} \,\ep(x)\, \Le D^{K} \Ra^{-1}\, \tilde{\ep}(\tilde{x}) - 
\,m_{p}^{3}\,\int d^4 x\,\sqrt{-g}\,\ep(x)
- \,m_{p}^{3}\,\int d^4 \tilde{x}\,\sqrt{-\tilde{g}}\,\tilde{\ep}(\tilde{x})\,.
\eeq
We see that there is no contribution to the cosmological constant from the $\ep$ fields in the action, they are quantum ones. 
The $\xi_{g}$ function here is proportional to the Keldysh propagator:
\beq\label{Worm9}
\xi_{g}(x,\,\tilde{x})\,\propto\,D^{K}(x,\,\tilde{x})\,=\,0
\eeq
which is zero after the proper regularization. Therefore, the first non-trivial contribution into the cosmological constant in the example will arise  only if we will 
introduce interaction potential between the $\ep$ fields. After the potential $V(\ep,\,\tilde{\ep})$ will be introduced, the constant will be equal to
the one-loop scalar self-energy contribution $\Sigma$. Again, we obtained that the non-zero value of the constant in the example is due the quantum effects.

 In the case when the matter fields will be included in the calculations, instead the
\eq{Worm1} expressionthe the gravity effective action will acquire the following form (we write it in the short simplified notations):
\beq\label{Worm102}
\Gamma_{w}\,=\,\sum_{k,\,l,\,m,\,n}\,\Le -g(x) \Ra^{k/2}\,\phi^{m}(x)\,V_{k\,l\,m\,n}\,\Le -g(\tilde{x}) \Ra^{l/2}\,\tilde{\phi}^{n}(\tilde{x})\,.
\eeq
The structure of the effective vertices in the \eq{Worm102} will be depend, therefore, on the present or absence of the  possible interactions of auxiliary fields with the fields of the matter.

\section{Extended solution and bimetric models}

 Interesting observation which we need to underline is that the proposed model in some operational or mathematical sense is very similar to the bimetric models widely applied in the alternative 
dark matter theories, see \cite{Petit1}. In particular, the framework we introduced is similar to the concept of weakly coupled worlds (WCW) introduced in \cite{DamKog}. 
Nevertheless, what is called as the weakly coupled worlds in \cite{DamKog}
in the present formulation are parts of the extended classical solution of the Einstein equations related by the CPTM transform.
Namely, in our model these worlds are different manifolds, A and B manifolds,
 which are glued on the quantum level whereas
the worlds of \cite{DamKog}  model "live" on the same  manifold interacting classically.
Whereas the frameworks are coincide in the limit of non-interacting worlds of the Multiverse, the model of \cite{DamKog}
supposes the local interaction term between the metrics of the separated manifolds, in the proposed model we consider the non-local quantum interactions between the 
parts of the extended manifold which can be reduced to the local one only after some averaging procedure.
The similar construction of the non-local interactions terms is known in QCD, see \cite{Our}, for the gravitation purposes the calculations were presented  in the \cite{LipatovG}. 
There is still a difference of course, the calculations of \cite{LipatovG} are performed in an assumption of high-energy kinematics in the process of interaction of gravitating objects,
but the framework can be adopted to the case of arbitrary interactions as well.

 On the classical level, if we will neglect the difference in the origin of the terms in the Lagrangian and equations of motion, the coupled  equations for the metric's parts in both
frameworks will coincide. In this extend the theories are equivalent, the \eq{Sc2} Lagrangian is the same as introduced in \cite{DamKog}. From the point of view of general 
interpretation of the additional metric's field, the present negative mass manifold is similar to the \cite{Hoss} proposal for the
anti-gravity particles framework but with the same important difference. In our model there is no place to the negative mass particle in our branch of the Universe, they populate the B manifold
of the extended solution. Additionally, concerning the \cite{DamKog} and \cite{Hoss} models, the important question is about the symmetries present in the models. The main idea
of the discrete \eq{Add1} CPTM transform is that it relates the two manifolds, each of which is invariant separately in respect to the separated connected subgroup of the full Poincare group
(see \cite{Souriau}) , in the way that the metric presents it's form after the transform. In this case there is no a common metric's diffeomorphism as in \cite{DamKog} but two separated groups of symmetry related by 
CPTM discrete transform.

\section{Conclusion}

 In this note we considered the possible application of the reversal extended CPTM symmetry of the extended space-time solutions of Einstein equations to the resolution of 
the cosmological constant problem. By construction, the proposed model can be considered as a variant of Multiverse with different signs of gravitational mass, charge, radial coordinates and time direction
in the separated parts of the extended space-time which are related by the CPTM transform. 
The immediate simplest consequences of the model is a zero value of the  vacuum energy density
and overall zero electrical charge on the classical level, see first Section of the note. 
In this extend the model is initially free from the problems of zero vacuum energy and baryon asymmetry, it describes maximally symmetrical Multiverse.  The model has some similarities to two-time direction models proposed for the solution of the Universe's 
low initial entropy value, see \cite{Saharov}, CPT symmetric Universe model considered in \cite{CPT} and models of \cite{Sym,Sym1} .   

  Discussing the general action of the theory we note that it remains trivial if we do not introduce an interaction between the parts of the extended solution, see
also \cite{Linde}.  On the classical level this
interaction must be zero if we do not require to change the classical Einstein equations for each section of the Multiverse. 
An immediate result of the introduction of the interaction term and gluing of the different manifolds by the gravitations is that in the each separated manifold arises 
a term which play role of the cosmological constant in the Einstein equations even in absence of other quantum fields. Reformulating it stays that there is a dynamic classical evolution of the metric  
of each manifold in the form of Einstein equations with cosmological constant caused by the mutual interaction between the separated manifolds through the gravity only. 
This interaction determines the classical topology of the separated manifolds and changes the value of the cosmological constant during the evolution. 
Important that constant's small and non-zero value is due it's non-classical origin, it is equal to zero at the classical level and small
due it's quantum origin.  The proposed resolution of the cosmological constant problem is different from the
considered in \cite{ClassWorm} therefore, in the present framework the cosmological constant is the result of the influence of the quantum vertices which "glue"
the different parts of the general manifold. The vertices, in turn, arise as a consequence of the \eq{Worm1}
expansion of the classical metric which form is dictated by CPTM symmetry, in the weak field approximation they are effective vertices of the interactions between the manifolds in the \eq{Worm101} effective action. 

 As examples of the non-local vertices which glue the different manifolds we considered two models. Each of them consists an auxiliary field which must be constructed by special way. Namely,
these fields must not to provide contributions into the cosmological constant on the classical level but only on the quantum one. In both examples we considered  doublets of the auxiliary fields,
in the first one we proposed a doublets of the fields with components related by the CPTM symmetry. In this case we stay with the two separated scalar fields 
defined on the different manifolds at the end. Therefore, due the CPTM symmetry requests from Section 1, the classical contributions of the fields into the constant is zero, the only non-trivial contributions will arise
if we will account the interaction between the traces of the metric's fluctuations and/or non-flat contributions in the corresponding propagator. 
Introducing the interactions between the auxiliary fields on the different manifolds we will obtain also a many-legs effective quantum vertices. In general the interactions between the auxiliary fields 
will provide the complex quantum effective vertices which glue A and B manifolds in the form of the effective action for the auxiliary fields. It is important also, that in this example we do not have to introduce the interactions between the auxiliary fields in general. If we will expand the regions of the integration in the corresponding integrals on both $x$ and $\tilde{x}$ coordinates using CPTM symmetry consequences, then the correlator which glue two manifolds will arise in the calculation. This "minimal" quantum wormhole between the manifolds will provide a non-zero contribution
to the constant at higher perturbation order in the calculations with the fluctuations above the classical metric introduced.

  Another example we considered is a construction
of the auxiliary fields on the base of Keldysh approach to the non-equilibrium processes in condensed matter physics. In this case our doublets are the pair of classical and quantum fields defined on the each manifold separately. The classical value of the field in the doublets is defined through the non-local interaction of the manifolds, see \eq{Worm7}. Proceeding, we will obtain again that the 
remained auxiliary fields in the auxiliary action are the quantum ones, they will not contribute into the cosmological constant. In this case the non-local 
gravitational interaction between the A and B manifolds will be provided 
only on the quantum level if we will introduce an interaction term
between the quantum auxiliary fields. Therefore, the corresponding contribution into the cosmological term will be expressed through
the self-energy diagram for the auxiliary scalar fields. As in the first example, it is quantum and therefore small, after the interaction between the auxiliary fields is introduced the many-legs quantum vertices between the manifolds will arise in the form of the auxiliary effective action as well.

  The number of the twins regions in the model depends on the
basic bare geometry. There are only two pairs of the regions in the Schwarzschild's extended solution and 
infinitely many in the Reissner-Nordtr\"{o}m extended soluiton of the classical equations for example. 
From this point of view, the cosmological constant depends on the basic geometry of the extended solution and forms and types of the proposed vertices-wormholes. The interesting task, therefore, is a direct calculation of the constant in \eq{Sc9} and/or \eq{Sc17} for the different geometries of interests. 
The properties of the  modes propagating through the proposed vertices  are also interesting, the "bridges" connect the 
manifolds with the different signs of the mass in. Therefore the problem of the stability of the vertices is different from the discussed in \cite{Thorne}.
Consequently there is an additional interesting question arises,  this is a  problem of the determination of the connected many-legs  vertices  geometries
and classical metrics requested for the calculations. Namely, the N separated vertices (wormholes) geometry exists and known, see \cite{NWorm} for example, but in general 
there is a need in the geometry of connected N ends vertices as well. As mentioned above, the solution of this problem requests a construction of the action for the interacting auxiliary fields.
So far it is not clear on the base of which principals and reasons these interactions must be introduced.

 The last remark is about the properties of \eq{Sc18}. The energy-momentum of the matter there, $T_{\phi}$, contains also the contributions from the classical values of the $\tilde{\phi}$
field. 
Through the graviton's exchange
processes we can therefore consider a semi-classical or quantum or both contributions of the negative mass matter "condensate" from an another part of the manifold to
our Universe trough the usual gravity interactions, see for example \cite{Dolg1}. 
This "condensate" interacts with the usual matter only by gravity force and, in principle,  can be considered as a possible source of the dark matter
in our part of the Universe. Additional source of these particles can be a some quantum tunneling of them through/by proposed complex vertices, it can be a very interesting problem to
investigate as well.

\end{document}